\documentclass[preprint]{iucr}

\usepackage{color,xcolor}
\usepackage{xspace}
\RequirePackage{graphicx}
\usepackage{epstopdf}
\usepackage{mathtools, amsmath, amsthm, amssymb, amsfonts,bm}
\usepackage[12pt]{moresize}
\usepackage{multirow}
\usepackage{enumitem}
\usepackage{url}
\usepackage{chemformula}

\usepackage{threeparttable}
\usepackage{mathtools,xparse,amsmath}

\graphicspath{{./figures/}}

\definecolor{tagcolor}{HTML}{21a889}

\definecolor{dgreen}{HTML}{008000}

\newcommand{\be}{\begin{enumerate}[wide, labelwidth=!, labelindent=0pt,
        label=\textbf{\textcolor{blue}{\arabic*}.}]}
    \newcommand{\bei}{\begin{enumerate}}
        \newcommand{\ee}{\end{enumerate}}
    \newcounter{saveenumi}

\newcommand{\fig}[1]{Fig.~\ref{fig:#1}}
\newcommand{\figs}[1]{Figs.~\ref{fig:#1}}
\newcommand{\sect}[1]{Section~\ref{sec:#1}}

\newcommand{\tabl}[1]{Table~\ref{table:#1}}

\newcommand{\nba}[1]{}

\newcommand{\pydr}{\textsc{pyDataRecognition}\xspace}

\makeatletter
\newcommand{\floatcaption}{%
    \ifx \@captype \@undefined \@latex@error {\noexpand \caption outside float}\@ehd \expandafter \@gobble \else \refstepcounter \@captype \expandafter \@firstofone \fi {\@dblarg {\@caption \@captype }}%
}%
\makeatother

\journalcode{A}              

\begin{document}                  



\title{Towards a machine-readable literature: finding relevant papers based on an uploaded powder diffraction pattern}
\shorttitle{pydatarecognition}

\author[a,\dagger]{Berrak}{Özer}
\author[b,\dagger]{Martin~A.}{Karlsen}
\author[a]{Zachary}{Thatcher}
\author[a]{Ling}{Lan}
\author[d]{Brian}{McMahon}
\author[d]{Peter~R.}{Strickland}
\author[d]{Simon~P.}{Westrip}
\author[d]{Koh~S.}{Sang}
\author[e]{David~G.}{Billing}
\author[f]{Dorthe~B.}{Ravnsbæk}
\cauthor[a,c]{Simon~J.~L.}{Billinge}{sb2896@columbia.edu}

\aff[a]{Department of Applied Physics and Applied Mathematics, Columbia University, \city{New York}, NY, 10027, \country{USA}}
\aff[b]{Department of Physics, Chemistry and Pharmacy, University of Southern Denmark, DK-5230 Odense M, \country{Denmark}}
\aff[c]{Condensed Matter Physics and Materials Science Department, Brookhaven National Laboratory, \city{Upton}, NY, 11973, \country{USA}}
\aff[d]{International Union of Crystallography, \city{Chester}, CH1 2HU, \country{UK}}
\aff[e]{School of Chemistry, University of the Witwatersrand, Private Bag 3, PO WITS, 2050 \country{South Africa}}
\aff[f]{Department of Chemistry, Aarhus University, DK-8000 Aarhus C, \country{Denmark}}
\aff[\dagger]{B. Özer and M. A. Karlsen contributed equally to this work}
\maketitle
\begin{center}
    \includegraphics[width=0.85\columnwidth]{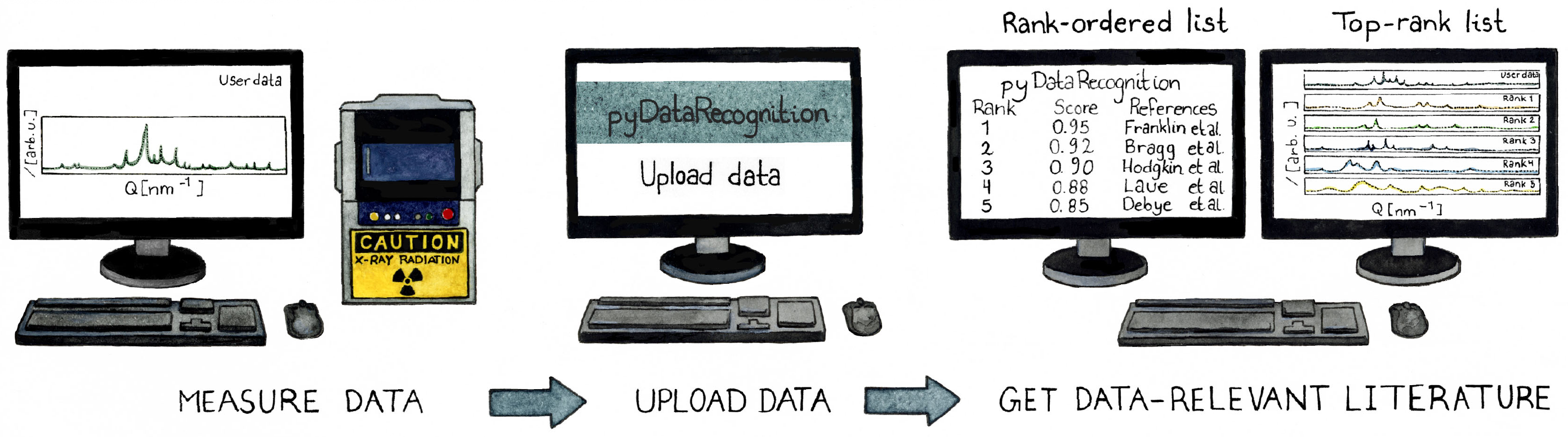}
    \label{fig:pydr_illustration}
\end{center}
\newpage
\begin{abstract}

We investigate a prototype application for machine-readable literature. The program is called \pydr and serves as an example of a data-driven literature search, where the literature search query is an experimental data-set provided by the user. The user uploads a powder pattern together with the radiation wavelength. The program compares the user data to a database of existing powder patterns associated with published papers and produces a rank ordered according to their similarity score. The program returns the digital object identifier (doi) and full reference of top ranked papers together with a stack plot of the user data alongside the top five database entries. The paper describes the approach and explores successes and challenges.

\end{abstract}
\section{Introduction}

The activity of communicating science, including paper writing, always includes a search of the literature to discover and acknowledge prior work \cite{garfieldWhenCite1996}.  Since the advent of the internet, this process has largely moved from manual,  library based, searches to online searches using search engines \cite{butlerSoupedupSearchEngines2000}.
Literature search engines such as Google Scholar \cite{vannoordenGoogleScholarPioneer2014} normally work by accepting text and metadata search queries, such as author names, keywords, journal name, year, and so on.  On the contrary, here we explore the concept of a data-seeded literature search where we use a measured data set as the search query to retrieve data-relevant papers from the literature.  We chose to use  X-ray powder diffraction data for our test case.

X-ray powder diffraction is an important technique in materials science, where structural characterization is at the very center of the workflow as it is inherently linked to material properties. The goal of the technique is to understand the arrangement of atoms in the material based on measurements of X-ray (or neutron or electron) diffraction.  When the sample is a powder, the resulting diffractogram is a one-dimensional (1D) pattern of peaks called a powder diffraction pattern \cite{gilmoreInternationalTablesCrystallography2019a,dinnebierPowderDiffractionTheory2008c}.
This serves as a 1D fingerprint of the structure of the material.

The challenge for our purposes is that there are no large open databases of experimental powder data.  The crystallography community recognized early the need for structured data related to chemical structure and developed the crystallographic information framework (CIF) \cite{hall;aca91}.  This was extended 
to allow for the capture of experimental data from powder experiments as part of the powder CIF development \cite{hallInternationalTablesCrystallography2006}. CIF dictionaries provide machine-readable definitions of data items that can appear in a CIF structured database such as a CIF file. Such CIF files (or 'CIFs') form the basis for multiple chemical (ICDD \cite{gates-rectorPowderDiffractionFile2019}, ICSD \cite{zagoracRecentDevelopmentsInorganic2019}, CSD \cite{groomCambridgeStructuralDatabase2016}, COD \cite{grazulisCrystallographyOpenDatabase2009d}), macromolecular (wwPDB \cite{bermanProteinDataBank2000}), and materials science (Materials Project \cite{jainCommentaryMaterialsProject2013d}) structural databases. 
However, of submitted CIFs that contain data resulting from a powder diffraction study, few include the associated diffractogram data (indeed, one desired outcome of this work would be to increase the incentives for authors to include the underlying diffractogram data).  The journals of the International Union of Crystallography (IUCr) archive all CIFs uploaded by authors with subsequently published papers. From this database we were able to extract a relatively small subset of CIFs that do contain powder patterns, along with metadata that allow the related paper to be found. This subset (787 CIFs) is the database we use for testing our prototype.

As a simple illustration of the kind of benefits that may derive from having a machine-readable capability for the contents of literature papers, we develop here a prototype application that would help experimentalists carry out a literature search early in the process of their study.  The specific use case that we want to demonstrate is described below, but the vision is a software  application that takes a measured powder pattern as input and returns a list of relevant papers in the existing literature, preferably without the user having to upload a significant amount of, or ideally any, additional information about the experiment. This is reminiscent of modern facial recognition capabilities but it is experimental data recognition and so we call the Python language based prototype application \pydr.

\section{Machine-readable vs. human-readable literature}

For the past $\sim 350$   years, the main goal of the scientific literature has been to condense scientific understanding into documents that are intelligible to humans.  It has been enormously successful by any measure.  However, the literature is growing rapidly and it is becoming difficult for humans to keep up with the number of new publications.  Also, it becomes hard to assimilate so much information and find correlations and insights between related studies.  When given structured data, machines are very good at finding correlations and clustering by similarity as exemplified by facial recognition algorithms \cite{anwarulComprehensiveReviewFace2020}.  There is an activity of using machine learning to have machines read papers that were written for humans to understand, a process called natural language processing (NLP) \cite{chowdharyNaturalLanguageProcessing2020}.  However, this process, whilst valuable for extracting information from the historic canon, is not the best way for machines to assimilate information from data.  We can expect much greater efficiencies in machine processing of the scientific information if we can take steps to make scientific papers readable by machines directly.

For this process to succeed, we need data in papers to be in accessible and structured data formats and saved with sufficient metadata to give important contextual information. The human being has a very highly developed capability for pattern recognition.  When we write a human-readable paper, we take our data and make an image, for example, by plotting the result as a line-plot.  It is much easier for the human reader to see similarities and derive insights from the data plotted as an image, but this is hard for the machine.  A literature that is written to be read by both humans and by machines would also have the data that was used to form the image saved in a machine-readable way, with important metadata such as what is being plotted, the quantity and units of both the $x$- and $y$-arrays, the sample that was measured to produce the plot, the people who did the work, and so on.  This is rarely done currently but is needed to realize the benefits of machine learned science.  

\section{Prototype literature search application}
In order to explore the kind of benefits that might be derived by having a machine-readable literature we explore a very simple use-case that makes use of a (small) database of structured, tagged, experimental data and does something useful with it.  The simple use-case we explore is to use a measured dataset as the input in a literature search.

\subsection{Use case}
\label{sec:uc}

The use-case is described in the following way.  A structure scientist has a powder diffraction pattern from a particular sample collected on their powder diffractometer. They upload the data to the search application, together with a limited amount of relevant experimental information. The application then will search a database of stored powder diffraction data associated with published papers. It will then return a list of relevant papers based on the similarity between the data uploaded by the user and the powder patterns appearing in the papers. In the simple first iteration of the concept the relevance will just be a ranking based on the similarity between the powder patterns in these papers and the powder diffraction data uploaded by the user.

The advantage of this use-case is that the International Union of Crystallography (IUCr) has a database of experimental powder patterns in a machine-readable powder-CIF format \cite{hallInternationalTablesCrystallography2006} that have been deposited by authors at the same time as they submitted the paper to the relevant IUCr journal.   These are the experimental data that generally appeared as images in figures in the linked papers.  The existence of this structured database of experimental powder patterns linked to published papers is therefore a valuable resource for prototyping the approach. 

We note that there are several databases that facilitate computational literature searches including CrossRef \cite{crossrefRESTAPI2020}, Scopus \cite{mongeonJournalCoverageWeb2016,burnhamScopusDatabaseReview2006}, Web of Science \cite{mongeonJournalCoverageWeb2016,mikkiGoogleScholarCompared2009}, arXiv \cite{ginspargArXiv202011}, Google Scholar \cite{mikkiGoogleScholarCompared2009, samadzadehComparisonFourSearch2013}, Google Image Search \cite{fergusLearningObjectCategories2005} and so on.  The purpose of this work is to show how properly tagged data held in a structured database can be included in literature search workflows, helping scientists to do better science more quickly.

\subsection{Software implementation}

The use case presented above has been implemented into a Python package. The package is using home-written functions based on well-established third-party libraries like NumPy \cite{harrisArrayProgrammingNumPy2020}, Matplotlib \cite{hunterMatplotlib2DGraphics2007a}, SciPy \cite{virtanenSciPyFundamentalAlgorithms2020b}, scikit-beam \cite{scikit-beamScikitbeam2022} to complete the use case.

To run the program, the user must provide the diffraction data for which the query should run. Currently, the data should be provided as a two-column  text file, possibly with a header, of intensity vs. an independent variable. The independent variable may be in the form of diffraction angle, $2\theta$ in degrees, $d$-spacing, in Ångströms, or the momentum transfer, $Q$, in inverse Ångströms, $\mathrm{\AA^{-1}}$. If the independent variable is $2\theta$,  the X-ray or neutron wavelength in Ångströms also needs to be provided.  All comparisons between data within the program are done with a $Q$ independent variable.  It will be straightforward to support different file formats in a production version of the code later.

The program then uses a distance metric to determine the similarity of the uploaded pattern to every pattern in the database.   In the current implementation we are using the Pearson correlation \cite{ pearsonVIINoteRegression1895}, $r_{xy}$,
\begin{equation}
    r_{xy} = \dfrac{\sum\limits_{i=1}^{n}(x_{i}-\overline{x})(y_{i}-\overline{y})}{\sqrt{\sum\limits_{i=1}^{n} (x_{i}-\overline{x})^{2}\sum\limits_{i=1}^{n}(y_{i}-\overline{y})^{2}}},
\end{equation}
where $x$ and $y$ are one-dimensional arrays of equal size and $\overline{x}$ and $\overline{y}$ are their means, respectively. The value of the correlation coefficient can vary between $+1$ and $-1$. A  value of $+1$ means the two data sets are identical (perfect positive correlation), 0 implies no correlation between the data sets. Numbers less than zero imply negative correlation. It is calculated using the \texttt{pearsonr()} method within the scipy.stats package \cite{virtanenSciPyFundamentalAlgorithms2020b}. Since our goal is to find similar data, we seek diffraction patterns with $r_{xy}$ close to one.

For a comparison of two data sets using the Pearson correlation, the two intensity arrays need to be on the same $Q$-grid. In general, powder patterns are measured over different ranges of $Q$ and on different arbitrary $Q$ grids.  To address this issue, we automatically determine the $Q$ space overlap region of the user and database data sets
and linearly interpolate the data onto a common regular $Q$ grid in this interval. 
Currently, a step size of $\Delta Q=10^{-3}\,\mathrm{\AA^{-1}}$ is used.  The user-supplied and target intensity arrays are then linearly interpolated onto this grid and the Pearson correlation is computed.  
Currently, the comparison is done over the full overlapping range as long as there is at least a 20~nm$^{-1}$ overlap.  If the overlap is smaller than this the database entry is not considered.  As a result of this heuristic, similarities are compared between pairs of data computed over different ranges of overlap. The Pearson measure is normalized by the number of points that are computed making comparisons between overlap regions of different length possible. This seems to give reasonable results but could be revisited in the future. 

The process of finding the overlapping range in $Q$ space, calculating a regular $Q$ grid, doing linear interpolation, and conducting Pearson correlation analysis between the user data and the data from a CIF is done for every CIF in the \pydr database. 
This is possible because of the small size of the database but will not scale to large databases of data and more efficient approaches will be investigated in the future.

The program then determines a rank ordered list based on similarity, and extracts from the database entry metadata the digital object identifier (DOI) \cite{paskinUniqueIdentifiers1999} of the paper that is associated with the the ranked data set.  The full reference of the associated paper is determined by making an API call to CrossRef \cite{crossrefRESTAPI2020} using the DOI.   The rank ordered list is then returned  to the user containing the rank, the Pearson-$r$ value, the DOI, and the full paper reference.  This information is also saved to a text file.

The five most similar \pydr database entries are plotted together with the user data to enable the user to visually inspect similarities between the data sets. 
Examples on output rank-ordered lists are given in Tables~\ref{table:rank_sandys1}-\ref{table:rank_sandys3} and plots in \figs{rank_plot_sandys1}-\ref{fig:rank_plot_sandys3} in Section~\ref{sec:results}.

\section{Outcomes}
\label{sec:outcomes}
\subsection{Results}
\label{sec:results}

At the moment of writing, $\sim515$ valid CIFs, out of 785 total in the \pydr database, are included in the analysis. They originate from $\sim215$ papers. The actual number of CIFs included in the analysis depends on the $Q$-range of the user data, as a minimum shared $Q$-range of 20~nm$^{-1}$ between the user and database patterns is required for the CIF to be included in the analysis.

Here, we explore the performance of the prototype \pydr with a number of query examples. The first example serves to test that the algorithm finds, with top rank, a dataset that actually exists in the database. The second example is a better test of the real use case. We provide real data but choose a very common structural form (perovskite) with the expectation that there will be representatives of this structure from more than one sample and composition, even given the  limited size of the current database of 785 CIFs. In the third example, we provide a neutron data set as user data to explore how the program behaves when provided neutron data as input whilst the data in the database come predominantly from X-ray data.

\textbf{Query example 1}

\begin{figure}
    \includegraphics[width=0.8\columnwidth]{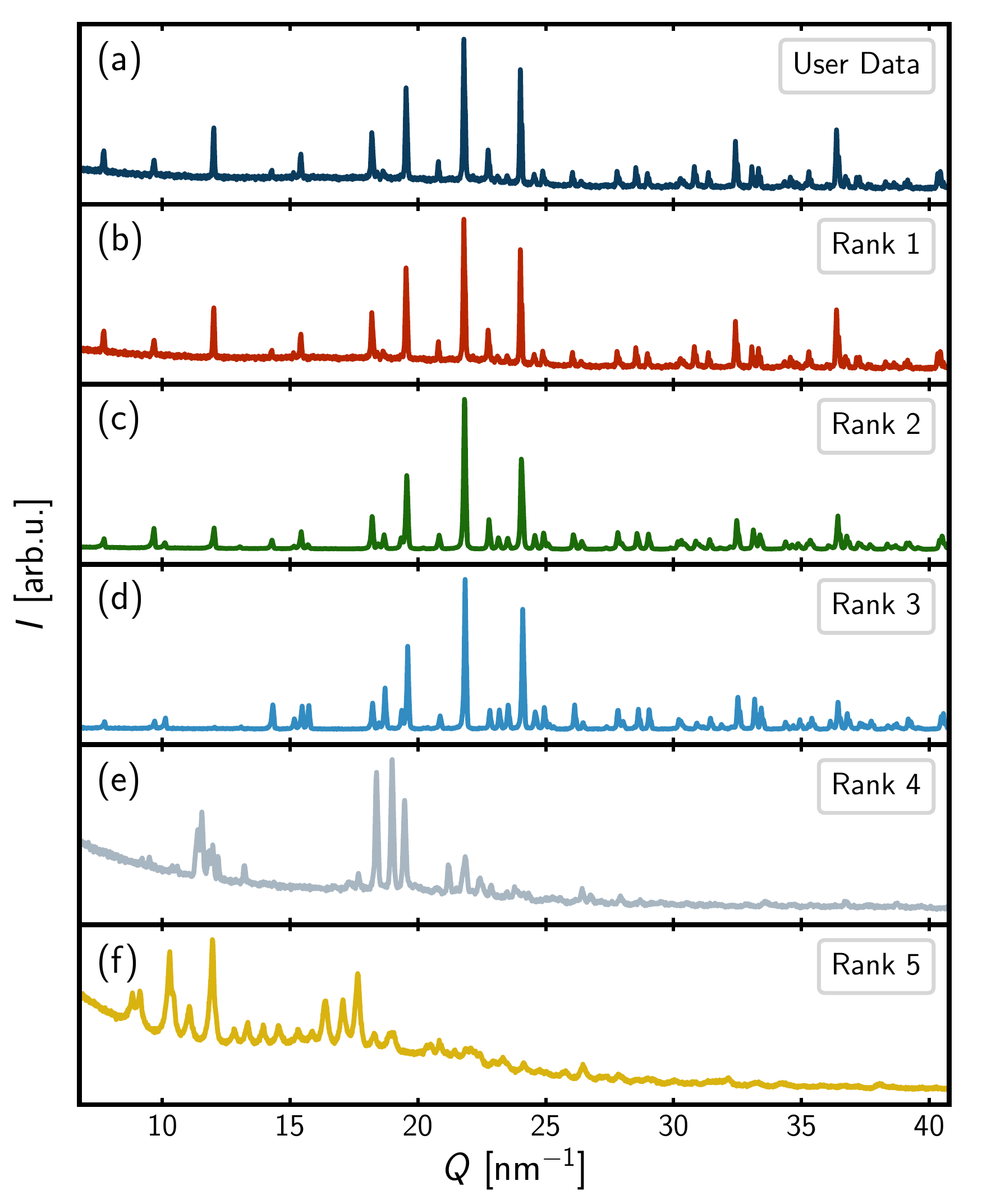}
    \label{fig:rank_plot_sandys1}
    \caption{Intensity, $I$, in arbitrary units as a function of momentum transfer, $Q$, in inverse nanometers, $\mathrm{nm}^{-1}$, for the user data (topmost) and the top five \pydr database entries in descending order.
    The full range of the user data are shown, whereas only the comparison region is shown for each database entry.
    The rank scores, DOIs, and references can be found in Table \ref{table:rank_sandys1}.}
\end{figure}

For the first example, the user data are taken from a CIF from the paper by \cite{stahliHydrogensubstitutedVtricalciumPhosphate2016}. The paper is on hydrogen-substituted $\beta$-tricalcium phosphate synthesized in organic media, i.e. a Mg-free whitlockite, represented by the
formula \ch{Ca_{21-x}(HPO4)_{2x}(PO4)_{14-2x}}, where $x = 0.80 \pm 0.04$. The data are from an X-ray experiment. 
As the user data are taken from a database entry, the expected outcome of the query is to have a perfect match, i.e. a score of one, $r_{xy}=1$. From \tabl{rank_sandys1}, 
it can be seen that the test went well, and a perfect match is found for a CIF appearing in the paper by \cite{stahliHydrogensubstitutedVtricalciumPhosphate2016}. 
\begin{table}
\caption{Ranks, scores, DOIs, and references for the top five \pydr database entries shown in Fig.~\ref{fig:rank_plot_sandys1}. The `user data' are identical to the rank 1 entry in the table.}
\begin{center}
\begin{threeparttable}
\begin{tabular}{c c c c}
    \textbf{Rank} & \textbf{Score} & \textbf{DOI} & \textbf{Reference} \\ \hline
      1 &   1.0000  &   https://doi.org/10.1107/S2052520616015675   & \cite{stahliHydrogensubstitutedVtricalciumPhosphate2016
} \\ \hline
      2 &   0.7379  &   https://doi.org/10.1107/S1600536810014327   & \cite{zatovskyRietveldRefinementWhitlockiterelated2010
} \\ \hline
      3 &   0.4631  &   https://doi.org/10.1107/S1600536813007848   & \cite{strutynskaRietveldRefinementAgCa102013
} \\ \hline
      4 &   0.4552  &   https://doi.org/10.1107/S2052520618004092   & \cite{bellCrystalStructuresK22018
} \\ \hline
      5 &   0.4261  &   https://doi.org/10.1107/S2052520614001140   & \cite{zvirgzdinsStructureDeterminationThree2014
} \\ \hline
\end{tabular}
\end{threeparttable}
\end{center}
\label{table:rank_sandys1}
\end{table}
In \fig{rank_plot_sandys1}(a) and (b), visual inspection confirms that the plots of the user data and the rank~1 database entry are identical. 
It is encouraging that the program returns the paper from which the user data were derived as the top ranked result. 

Moving down in \tabl{rank_sandys1}, the rank~2 entry \cite{zatovskyRietveldRefinementWhitlockiterelated2010}, which studies Rietveld refinement of whitlockite-related \ch{K_{0.8}Ca_{9.8}Fe_{0.2}(PO4)7}, scores 0.7379. The score indicates an intermediate level of similarity to the user data.  Visual inspection of the plot in \fig{rank_plot_sandys1}(c) confirms the similarity and the structural relation between the user data and the rank~2 entry that are both whitlockite-related. Multiple Bragg positions are shared between the two data sets, as reflected in the intermediate score, but at the same time dissimilarities are also present, such as differences in relative intensities and peak splitting, e.g. right above 10~nm$^{-1}$, as should be expected from different chemical compositions. The data-driven nature of the \pydr query enables the user to discover other papers with possible relevance to their uploaded data. 

The rank~3 dataset has a much lower agreement factor (0.4631), than the rank~2 (0.7379) which might suggest that it is structurally unrelated.  However, visual comparison of the diffraction curves (\fig{rank_plot_sandys1}(c) and (d)) suggests that there are many similarities between these datasets.  In fact, The rank~3 dataset \cite{strutynskaRietveldRefinementAgCa102013} is from a Rietveld refinement study of  a sample isostructural to the mineral whitlockite, \ch{AgCa10(PO4)7}, which is closely related to the user dataset and would certainly be of interest to the user. In this case the Pearson measure seems not ideal as a similarity metric for the current use-case.

\begin{figure}
    \includegraphics[width=0.8\columnwidth]{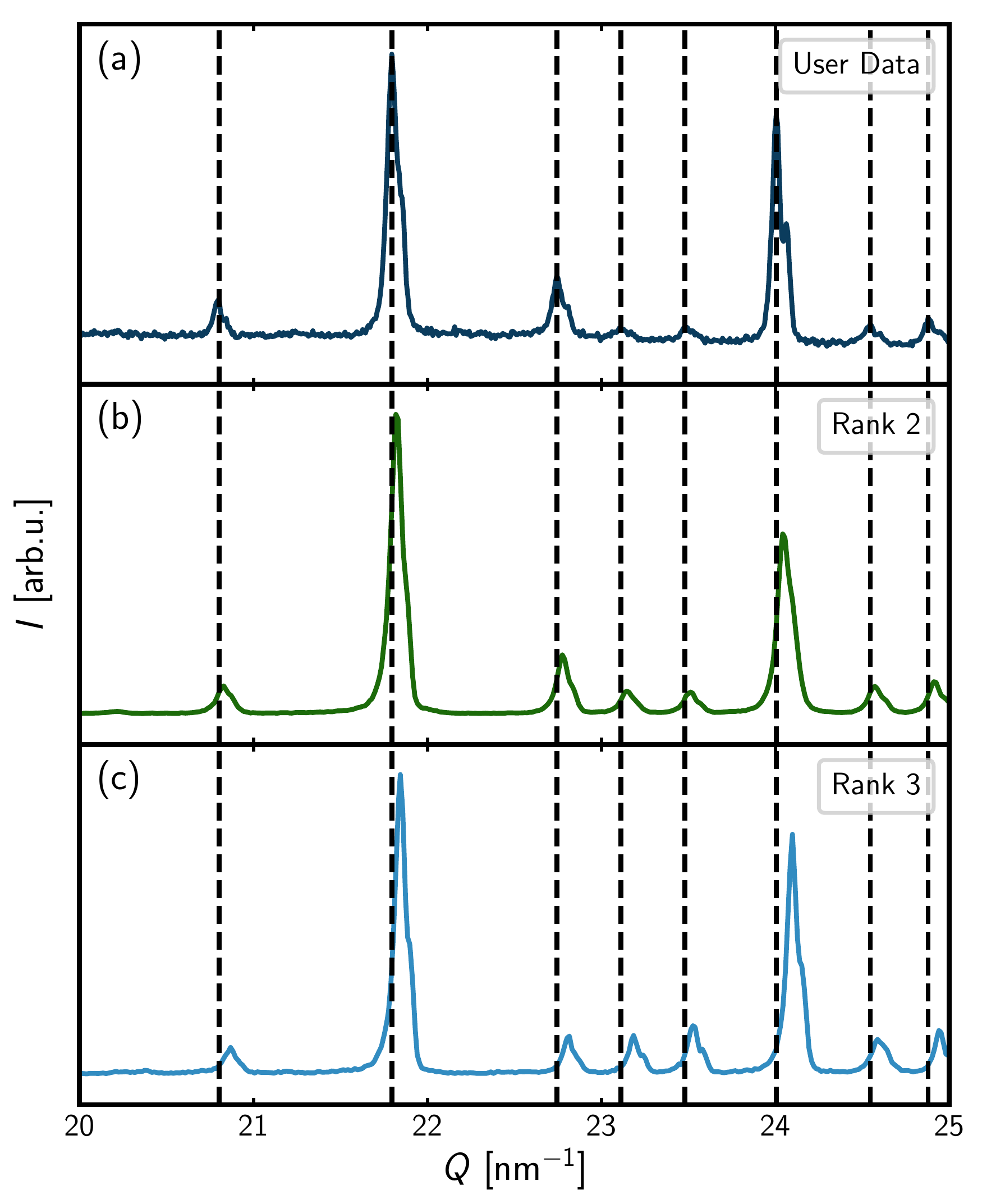}
    \label{fig:rank_plot_sandys1_zoom}
    \caption{Intensity, $I$, in arbitrary units as a function of momentum transfer, $Q$, in inverse nanometers, $\mathrm{nm}^{-1}$, for the user data (topmost) and the rank~2 and 3 \pydr database entries.
    The data are plotted for the $Q$-range from 20 to 25~nm$^{-1}$.
    The vertical lines indicate the Bragg positions of the user data.}
\end{figure}

To explore the origin of the large drop in similarity score between the isostructural rank~2 and~3 samples in \fig{rank_plot_sandys1_zoom} we have plotted the user data together with the rank~2 and~3 database entries on an expanded $Q$-scale from~20 to 25~nm$^{-1}$ with vertical lines indicating the peak positions of the user data. 
We see that there is a small offset in peak position for the rank~2 database entry relative to that of the user data, whereas the offset is more pronounced for the rank~3 database entry. This $Q$-offset is likely to explain the low Pearson score of the rank~3 entry. The difference in scores for the rank~2 and~3 database entries gives a hint at how sensitive the current Pearson similarity metric is towards an offset, whether it is an experimental artefact or has a structural origin such as different lattice parameters of otherwise similar structures.
This is undesirable behavior in our similarity metric that we explore further below. 

The rank~4 and~5 entries in \tabl{rank_sandys1} and \fig{rank_plot_sandys1}(e-f) in the current example appear visually very dissimilar to the user data and are unlikely to be of any interest to the user. However, it is observed that the Pearson scores are quite similar to that of the rank~3 entry that is isostructural.  This is further evidence of a weakness in the use of the Pearson metric in the current application as it cannot distinguish an isostructural but shifted pattern from a completely dissimilar pattern.  The code was designed for it to be easy to implement different similarity metrics in principle, and finding the best similarity metrics will be an ongoing process.

\textbf{Query example 2}

\begin{figure}
    \includegraphics[width=0.8\columnwidth]{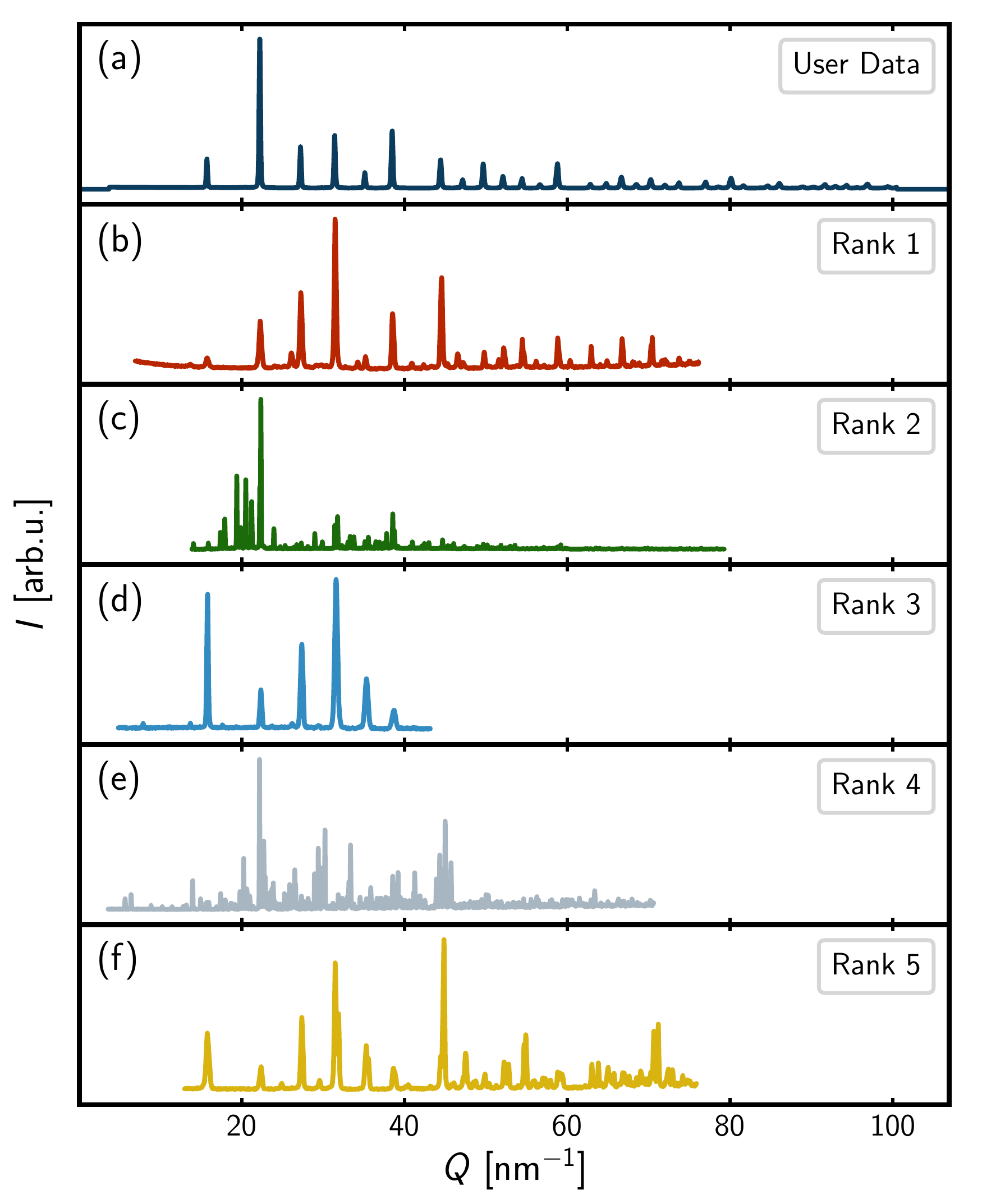}
    \label{fig:rank_plot_batio3}
    \caption{Intensity, $I$, in arbitrary units, arb.u., as a function of momentum transfer, $Q$, in inverse nanometers, $\mathrm{nm}^{-1}$, for the user data (topmost) and the top five \pydr database entries in descending order.
    The full range of the user data is shown, whereas only the region used for comparison is shown for each database entry.
    The rank scores, DOIs, and references can be found in Table \ref{table:rank_batio3}.}
\end{figure}

For the second example, the input data are synchrotron X-ray data of the perovskite \ch{BaTiO3}. Since the structural family of perovskites is common, it is hoped that even the current small database will return one or more papers with data from perovskite or perovskite-related structures. The results from the query are found in \tabl{rank_batio3} and \fig{rank_plot_batio3}. 
\begin{table}
\caption{Ranks, scores, DOIs, and references for the top five \pydr database entries shown in \fig{rank_plot_batio3}.}
\begin{center}
\begin{threeparttable}
\begin{tabular}{c c c c}
    \textbf{Rank} & \textbf{Score} & \textbf{DOI} & \textbf{Reference} \\ \hline
      1 &   0.5723  &  https://doi.org/10.1107/S0021889813013253   & \cite{iturbe-zabaloSymmetrymodeAnalysisPhase2013}  \\ \hline
      2 &   0.3906  &  https://doi.org/10.1107/S0108768198017984  & \cite{sciauStructuresPhasesParaelectrique1999}  \\ \hline
      3 &   0.3390  &  https://doi.org/10.1107/S1600576715000941   & \cite{orayechModecrystallographyAnalysisCrystal2015}  \\ \hline
      4 &   0.2881  &  https://doi.org/10.1107/S0108768111039759   & \cite{kasunicStructureLaTi2Al9O19Reanalysis2011}  \\ \hline
      5 &   0.2485  &  https://doi.org/10.1107/S0108768109011057   & \cite{zhangStructuresK005Na02009}  \\ \hline
\end{tabular}
\end{threeparttable}
\end{center}
\label{table:rank_batio3}
\end{table}
From the scores reported in \tabl{rank_batio3}, it is evident that no highly similar database entries are encountered as all scores $r_{xy}<0.6$.  However, a visual inspection of the top ranked powder pattern in \fig{rank_plot_batio3}(b), does show some similarity in peak frequency and positions, so the rank~1 entry may be related to the user data despite the modest score of 0.5723 reported in \tabl{rank_batio3}. Looking into the paper  \cite{iturbe-zabaloSymmetrymodeAnalysisPhase2013}, the topic is symmetry-mode analysis of the phase transitions in \ch{SrLaZnRuO6} and \ch{SrLaMgRuO6} ordered double perovskites, i.e. a paper on perovskite-derived structures, which is encouraging, considering that the user data were for the perovskite \ch{BaTiO3}, so from all of the $>500$ entries in the database, \pydr has returned a paper describing related data in the top rank position, albeit with a low similarity score. 

Returning to the remaining results reported in \tabl{rank_batio3}, it is seen that all scores are $<0.4$, indicating low Pearson similarity to the user data. For the rank~2 and~4 entries, the low scores seem to reflect structural dissimilarity as the diffraction patterns are visually very different. The rank~2 entry \cite{sciauStructuresPhasesParaelectrique1999} is considering the structures of the paraelectric and ferroelectric phases of \ch{Pb2KNb5O15} with orthorhombic symmetry that does appear to be perovskite-related. \fig{rank_plot_batio3}(c) shows that the database entry possess a much larger peak density compared to the user data, as also reflected in the rather low score of 0.3906.

The paper of the rank~4 entry \cite{kasunicStructureLaTi2Al9O19Reanalysis2011} is on the structure of \ch{LaTi2Al9O19}, a non-perovskite compound isostructural to \ch{SrTi3Al8O19}, and so the low score of 0.2881 again reflects a structural dissimilarity.
However, the low Pearson score for the rank~3 and rank~5 results seem surprising, as in these cases the data have a  visual resemblance to the user data in \fig{rank_plot_batio3}(a), (d) and (f), especially in the rank~3 case. The paper of the rank~3 entry \cite{orayechModecrystallographyAnalysisCrystal2015}, is considering mode-crystallography analysis of the crystal structures and the low- and high-temperature phase transitions in the \ch{Na_{$0.5$}K_{$0.5$}NbO3} cubic perovskite. This paper clearly describes a closely related structure and we would hope that the \pydr algorithm would find it with a high ranking yet it does not.  In the case of the rank~5 entry \cite{zhangStructuresK005Na02009} it does also describe perovskite structures (\ch{K_{$0.05$}Na_{$0.95$}NbO3} and \ch{K_{$0.30$}Na_{0.70}NbO3}).

As was the case for the first query example, the low scores of the otherwise visually similar rank~3 and~5 entries may be explained by a small offset in the lattice parameters. For both entries, the offset is towards higher $Q$-values, compared to the user data, which is the likely cause of the poor Pearson score. In addition, there are also clear differences in relative peak intensities compared to the user data and some peak broadening, e.g., at 32~nm$^{-1}$, which will also affect the Pearson score.

The rank~3 and~5 entries represent additional false negative results.  These results did show up in the top-5 list despite their low Pearson scores, which is encouraging, but it is likely that other related papers are being missed with low Pearson scores because of the poor performance of Pearson for the job in hand.

\textbf{Query example 3}

\begin{figure}
    \includegraphics[width=0.8\columnwidth]{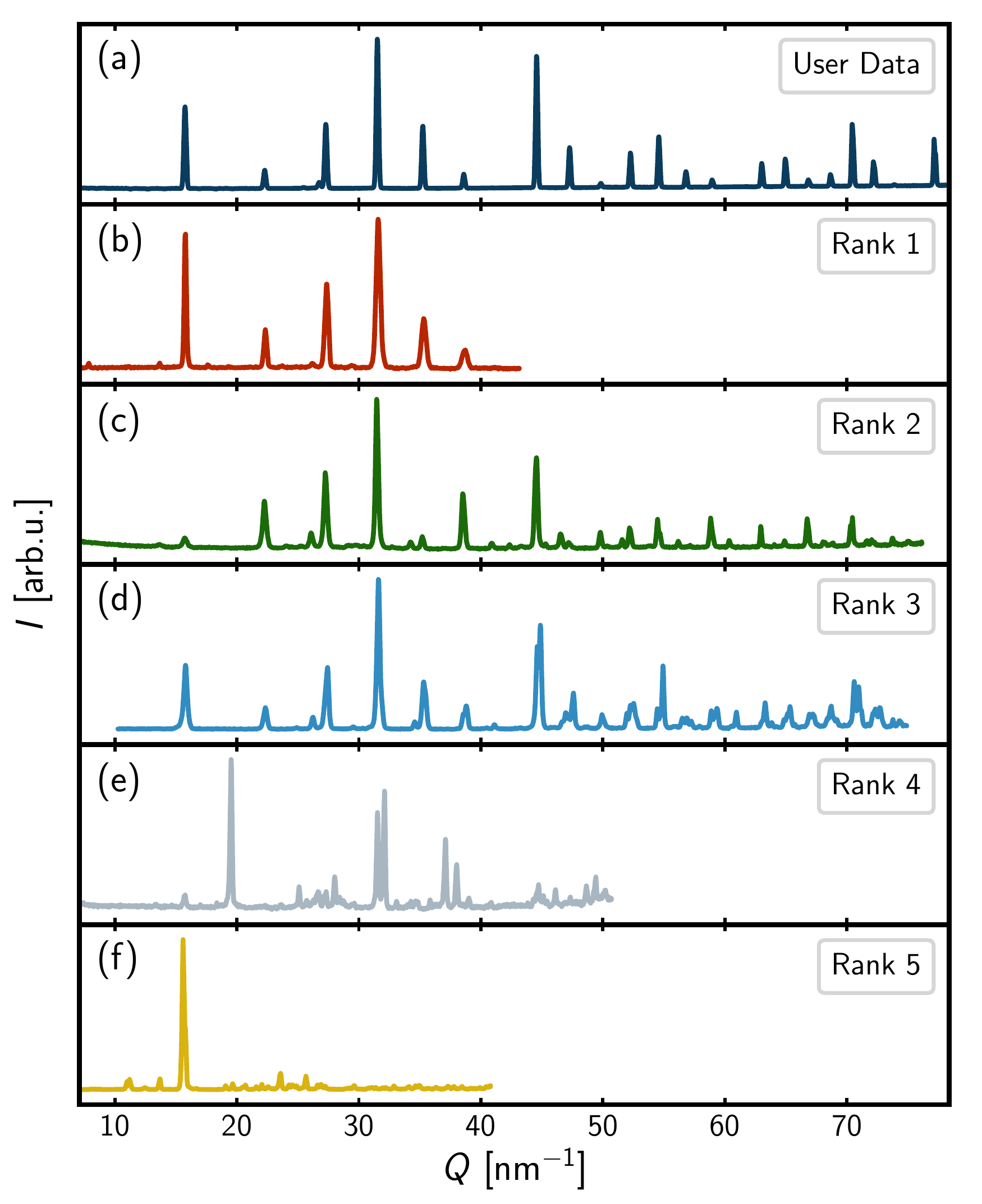}
    \label{fig:rank_plot_sandys3}
    \caption{Intensity, $I$, in arbitrary units, arb.u., as a function of momentum transfer, $Q$, in inverse nanometers, $\mathrm{nm}^{-1}$, for the user data (topmost) and the top five \pydr database entries in descending order.
    The full range of the user data are shown, whereas only the region used for comparison is shown for each database entry.
    The rank scores, DOIs, and references can be found in \tabl{rank_sandys3}.}
\end{figure}

The third and last query example reported here is regarding user data for which neutrons were used as the probe, contrary to the two former query examples that originated from X-ray probes. \pydr accepts powder patterns from any source, X-ray, neutron, or electrons, and currently the user is not asked to provide the type of probe on input, just as the type of probe is not regarded when running the query. Regardless, in principle we would still like the program to return papers describing similar structures. The results are shown in \tabl{rank_sandys3} and \fig{rank_plot_sandys3}. 
\begin{table}
\caption{Ranks, scores, DOIs, and references for the top five \pydr database entries shown in \fig{rank_plot_sandys3}.}
\begin{center}
\begin{threeparttable}
\begin{tabular}{c c c c}
    \textbf{Rank} & \textbf{Score} & \textbf{DOI} & \textbf{Reference} \\ \hline
      1 &   0.8808  &  https://doi.org/10.1107/S1600576715000941   & \cite{orayechModecrystallographyAnalysisCrystal2015}  \\ \hline
      2 &   0.7785  &  https://doi.org/10.1107/S0021889813013253   & \cite{iturbe-zabaloSymmetrymodeAnalysisPhase2013}  \\ \hline
      3 &   0.6855  &  https://doi.org/10.1107/S0108768109011057   & \cite{zhangStructuresK005Na02009}  \\ \hline
      4 &   0.2859  &  https://doi.org/10.1107/S0108768112017478   & \cite{bereciartuaStructureRefinementSuperspace2012c}  \\ \hline
      5 &   0.2532  &  https://doi.org/10.1107/S0108768103019013   & \cite{palaciosPhasesCH34N2003}  \\ \hline
\end{tabular}
\end{threeparttable}
\end{center}
\label{table:rank_sandys3}
\end{table}
The user data set is measured for a cubic perovskite sample, \ch{(K_{0.48}Na_{0.48}Li_{0.04})Nb_{0.98}Mn_{0.02}O3} \cite{h.e.mgbemereElectricFieldInducedPhaseTransition2017}, like the second query example above. From \tabl{rank_sandys3}, a relatively high score of 0.8808 is obtained for the rank~1 database entry \cite{orayechModecrystallographyAnalysisCrystal2015}. 
The paper for this entry is on \ch{Na_{$0.5$}K_{$0.5$}NbO3} perovskite and reports neutron data, explaining the high degree of similarity.
The rank~2 \cite{iturbe-zabaloSymmetrymodeAnalysisPhase2013} and~3 \cite{zhangStructuresK005Na02009} database entries  have slightly lower Pearson coefficients (0.7785 and 0.6855, respectively) and the visual similarity between the two is evident, though the rank~3 dataset is observed to have lower visual similarity to the user dataset. Mostly peaks are in the same place but relative intensities are quite different and peaks are split. The rank~2 database entry was for neutron diffraction data of the double perovskites, \ch{SrLaZnRuO6} and \ch{SrLaMgRuO6}, encountered before, and indeed weak additional superlattice peaks from the ordering are evident in the pattern.  The rank~3 database entry  describes neutron data precisely from perovskite K/Na niobate materials, like the user data, albeit with different K:Na ratios, \ch{K_{$0.05$}Na_{$0.95$}NbO3} and \ch{K_{$0.30$}Na_{$0.70$}NbO3} as well as the absence of Mn. In this case, peak splitting indicates a symmetry lowering in the database data, but it was correctly detected as closely related by the Pearson similarity in this case.
The rank~3 paper \cite{zhangStructuresK005Na02009} was also encountered as the rank~5 entry in \tabl{rank_batio3} of the second query example. However, the data plotted in \fig{rank_plot_batio3}(f) stems from a CIF considering \ch{K_{$0.30$}Na_{$0.70$}NbO3} at 200~K, whereas the data plotted in \fig{rank_plot_sandys3}(d) stems from a CIF considering the material at 180~K and the temperature difference may explain the slight visual differences when comparing the two database entries to one another.

Finally, the very low scores of the rank~4 and~5 entries in \tabl{rank_sandys3} are reflected by the observed dissimilarity to the user data in \fig{rank_plot_sandys3}(e) and (f) and in this case is due to a structural dissimilarity. The rank~4 entry \cite{bereciartuaStructureRefinementSuperspace2012c} describes the system
\ch{Bi_{$2(n+2)$}Mo_{$n$}O_{$6(n+1)$}} ($n=3$, 4, 5, 6) and the rank~5 entry \cite{palaciosPhasesCH34N2003} \ch{[(CH3)4N](ClO4)} at low temperature, neither of which are perovskite related structures.

Overall, the approach is working, with \pydr successfully suggesting to the user, from a database of 785 datasets (of which $\sim 600$ are usable) the same three perovskite structures in comparison to user inputs from perovskite structures, working for both X-ray data from \ch{BaTiO3} and neutron data from a perovskite Na/K niobate.   However, the test revealed certain difficulties that the Pearson correlation coefficient was having at correctly identifying and ranking nearby structures, especially when there was a small shift in the peaks due to different lattice parameters.

\subsection{Challenges and opportunities}
\label{sec:challengesandopportunities}

The completed use case demonstrates a proof of concept and reveals the great potential of a machine-readable literature.  There is still some way to go before it becomes a practical tool but the prototype highlights some of the challenges as well as the opportunities.

Currently the biggest limitation we encounter is the small database size. Of all the CIFs in the IUCr database (currently numbering around 100,000) only $\sim 1000$  contained experimental powder patterns. In part this is because many studies did not involve the use of powder diffraction data, but also there is limited adoption by authors of the ability to store the actual powder data.  Although, through the CIF mechanism, the IUCr is a leader in capturing the powder diffraction data of authors in a structured way, the uptake of the community is still limited.  This is currently a focus of the Commission on Powder Diffraction of the IUCr, where new tools for validating deposited CIFs for contained powder data, and tools for visualizing deposited data are being developed.  This \pydr prototype application adds additional incentive to authors as it will clearly make their work more discoverable in the future, and it is only a first step of what can be done if structured data are stored along with the papers describing them.

The current similarity metric, the Pearson correlation coefficient, is a good first step as a similarity measure that can easily be implemented in this prototype application. However, \sect{outcomes} illustrates that the job currently done by the Pearson correlation does not completely meet the requirements of the \pydr program, as it is observed that the Pearson correlation seems quite sensitive to $Q$ offset. This results in rather low ranking of otherwise visually similar patterns, which is undesirable from a user's point of view.
For experimental data, experimental artefacts from the instrument and the sample are expected, including e.g. $Q$ offset. For \pydr, it is desirable to have a similarity metric that tolerates the presence of experimental artefacts and still ranks otherwise similar patterns high to each other.
Apart from $Q$ offset, potential experimental artefacts to be tolerated include e.g. peak broadening from the instrument as well as sample, small peak splittings indicating slight losses in symmetry, and to some degree, variations in relative intensities of peaks as expected when comparing neutron to X-ray data, or from isostructural but different composition samples that may still be relevant to the user.
A better similarity metric for the current job required by \pydr would tolerate these aberrations and ideally return results by relevance, much like a Google search does, given the right search query.  We will explore different metrics, including ones specifically proposed for powder data 
\cite{degelderGeneralizedExpressionSimilarity2001}, but this is a big area of research \cite{vombrockeStandingShouldersGiants2015} and extending the metric is beyond the scope of the current article.

The prototype also highlights another challenge, which is maintaining the quality of the deposited data and the attached metadata.
Despite the small database size, a significant number of the deposited CIFs containing measured data were unusable.
Out of 787 CIFs, 785 could be parsed using the CIF parser \texttt{CifFile.ReadCif()} from the pyCIFRW Python module \cite{hesterValidatingCIFParser2006}.
Of those parsed, multiple CIF keys had to be browsed for $2\theta$, intensity, and wavelength values, the main reason being that CIF handles both measured, processed, and calculated data. The pdCIF dictionary makes it easier for developers to find the right keys, at least if the use of the keys follows the pdCIF guidelines. \cite{tobyPowderDictionaryPdCIF2006} However, from the current work, this does not seem to be the case in many instances, the result being that more CIF keys have to be browsed than if the CIF contributors strictly followed the pdCIF guidelines from IUCr. It may be beneficial to demand CIF contributors to obey the pdCIF guidelines, as this will reduce the number of keys to be browsed in the light of any machine readable literature effort.

For 59 of the 785 CIFs, the wavelength was missing, preventing a conversion to a physics-based independent variable such as $Q$. Furthermore, for 164 CIFs, the $2\theta$-values were either not stated explicitly or could not be calculated using a CIF supplied minimum, maximum, and step-size in $2\theta$.
Whilst it might be possible to modify the algorithm to guess at a resolution for the inconsistent min-max-step calculation, for example, by ignoring the author supplied bin-size but computing it from the minimum and maximum $2\theta$ values and the number of entries in the intensity array, this is not preferred behavior as it is modifying the user data in possibly ambiguous ways, and so in these cases the CIFs were discarded.
Inconsistent size of the min-max-step calculated $2\theta$ array relative to that of the intensity array was encountered for 92 out of all the 164 CIFs. A min-max-step calculated $2\theta$ array consistent with the size of the intensity array was obtained for 120 out of the 785 CIFs processed. None of the CIFs in the current database had $x$-axis data stored in $Q$ or $d$ quantity, though this would be supported if encountered.

These challenges highlight the need for better validation of CIF inputs of experimental data, as well as more intuitive and easy-to-use tools for experimenters to upload their experimental data and provide the needed metadata.

\subsection{Next steps}

The work described here is an early prototype for data-driven literature search to illustrate the potential for the machine-readable literature concept. It has served to illustrate the concept and to explore what some of the challenges will be to bring this to fruition.  An obvious next step would be to increase the size of the database. We are working with members of the Commission on Powder Diffraction at the IUCr to find ways to increase the amount and readability of powder diffraction data being deposited with submitted papers. In the mean-time one approach to increasing the size of the database is to simulate powder patterns from all structures (including those solved from single-crystal data) in the IUCr CIF database. Comparisons could then be made between uploaded data and simulated, as well as experimental, patterns. 

Another issue is finding similarity between diffraction patterns that were measured from the same material but measured under different experimental conditions of instrument resolution and so on.  It will be interesting to explore using different deconvolution methods and different representations for the data to see which are most effective.

For the current database consisting of 785 CIFs, on a laptop (HP Elitebook 850 G5, Intel(R) Core(TM) i5-7300 CPU @ 2.6 GHz, 2712 MHz, 2 cores, 4 logic processors, 8 GB RAM), it takes the program $\sim30$~s to complete a query.  This is acceptable, but will not scale well with larger datasets.

We currently use a  \textit{brute force} approach for finding similarity which will not scale well requiring more sophisticated and faster database browsing approaches to be found.
Prior information from the user can help, for example, a list of chemical elements that the user knows should be present in the sample, would cut down the search space.  However, we will also explore increasing the efficiency or the search, for example, through the
use of graph-based search algorithms that can pre-store the similarity between every entry in the database. \cite{johnsonBillionScaleSimilaritySearch2021}. Algorithms for finding nearest neighbor connections may then be explored to rapidly find the best solutions without having to traverse the entire graph.

\section{Conclusions}

As a first step towards a more machine-readable literature that will ease literature search and make science more readily available, we have demonstrated a prototype application, \pydr. The program takes a measured powder pattern, together with other relevant metadata, as input and returns information on literature papers that may be relevant to the powder pattern uploaded by the user.

This represents the initial steps towards a more machine-readable literature.  However, it already revealed a number of challenges that need to be overcome moving forward. The CIF format is well-defined but is not strictly adhered to or validated, at least when it comes to experimental data in powder-cif.  This results in non-usable information in the CIF file database such as non-numeric values where numeric values are expected. Tools are needed to facilitate the deposition of properly validated data-containing CIF entries in the IUCr database.  This work is in progress. Regardless, the simple use-case of finding relevant papers given a diffraction pattern already gives a glimpse of many other more advanced capabilities that are possible by going down this route of a machine-readable literature.

\section{Acknowledgements}

We would like to thank Dr. Sandra Skjærvø for sharing unpublished \ch{BaTiO3} data.

\section{Funding information}

Work in the Billinge group was funded in part by the U.S. National Science Foundation through grant DMREF-1922234. M. A. Karlsen and D. B. Ravnsbæk acknowledge the support from the Carlsberg Foundation (grant. no. CF17-0823). 
\newpage
\bibliographystyle{iucr}
\bibliography{billinge-group-bib,bg-pdf-standards,bo_pydatarecognition}

@article{pearsonVIINoteRegression1895,
	title = {{VII}. {Note} on regression and inheritance in the case of two parents},
	volume = {58},
	url = {https://royalsocietypublishing.org/doi/abs/10.1098/rspl.1895.0041},
	doi = {10.1098/rspl.1895.0041},
	number = {347-352},
	urldate = {2022-01-18},
	journal = {Proceedings of the Royal Society of London},
	author = {Pearson, Karl and Galton, Francis},
	month = jan,
	year = {1895},
	note = {Publisher: Royal Society},
	pages = {240--242},
}

@misc{scikit-beamScikitbeam2022,
	title = {scikit-beam},
	url = {https://github.com/scikit-beam/scikit-beam},
	urldate = {2022-01-27},
	publisher = {scikit-beam},
	author = {scikit-beam},
	month = jan,
	year = {2022},
	note = {original-date: 2014-07-10T04:44:35Z},
}

@article{hall;aca91,
	title = {The crystallographic information file ({CIF}): a new standard archive file for crystallography},
	volume = {47},
	copyright = {Copyright (c) 1991 International Union of Crystallography},
	issn = {0108-7673},
	shorttitle = {The crystallographic information file ({CIF})},
	url = {//scripts.iucr.org/cgi-bin/paper?es0164},
	doi = {10.1107/S010876739101067X},
	language = {en},
	number = {6},
	urldate = {2022-02-10},
	journal = {Acta Cryst A},
	author = {Hall, S. R. and Allen, F. H. and Brown, I. D.},
	month = nov,
	year = {1991},
	note = {Number: 6
Publisher: International Union of Crystallography},
	pages = {655--685},
}

@article{anwarulComprehensiveReviewFace2020,
  title = {A {{Comprehensive Review}} on {{Face Recognition Methods}} and {{Factors Affecting Facial Recognition Accuracy}}},
  author = {Anwarul, Shahina and Dahiya, Susheela},
  year = {2020},
  journal = {Proceedings of ICRIC 2019},
  pages = {495--514},
  publisher = {{Springer, Cham}},
  doi = {10.1007/978-3-030-29407-6_36},
  langid = {english}
}

@article{bellCrystalStructuresK22018,
  title = {Crystal Structures of {{K2}}[{{XSi5O12}}] ({{X}} = {{Fe2}}+, {{Co}}, {{Zn}}) and {{Rb2}}[{{XSi5O12}}] ({{X}} = {{Mn}}) Leucites: Comparison of Monoclinic {{P21}}/c and {{Ia}}\{\textbackslash overline 3\}d Polymorph Structures and Inverse Relationship between Tetrahedral Cation ({{T}} = {{Si}} and {{X}})\textemdash{{O}} Bond Distances and Intertetra\-hedral {{T}}\textemdash{{O}}\textemdash{{T}} Angles},
  shorttitle = {Crystal Structures of {{K2}}[{{XSi5O12}}] ({{X}} = {{Fe2}}+, {{Co}}, {{Zn}}) and {{Rb2}}[{{XSi5O12}}] ({{X}} = {{Mn}}) Leucites},
  author = {Bell, A. M. T. and Henderson, C. M. B.},
  year = {2018},
  month = jun,
  journal = {Acta Crystallographica Section B: Structural Science, Crystal Engineering and Materials},
  volume = {74},
  number = {3},
  pages = {274--286},
  publisher = {{International Union of Crystallography}},
  issn = {2052-5206},
  doi = {10.1107/S2052520618004092},
  copyright = {Copyright (c) 2018 International Union of Crystallography},
  langid = {english}
}

@article{bereciartuaStructureRefinementSuperspace2012c,
  title = {Structure Refinement and Superspace Description of the System {{Bi2}}(n + 2){{MonO6}}(n + 1) (n = 3, 4, 5 and 6)},
  author = {Bereciartua, P. J. and Zu{\~n}iga, F. J. and {Perez-Mato}, J. M. and Pet{\v r}{\'i}{\v c}ek, V. and Vila, E. and Castro, A. and {Rodr{\'i}guez-Carvajal}, J. and Doyle, S.},
  year = {2012},
  month = aug,
  journal = {Acta Crystallographica Section B: Structural Science},
  volume = {68},
  number = {4},
  pages = {323--340},
  publisher = {{International Union of Crystallography}},
  issn = {0108-7681},
  doi = {10.1107/S0108768112017478},
  copyright = {Copyright (c) 2012 International Union of Crystallography},
  langid = {english}
}

@article{bermanProteinDataBank2000,
  title = {The {{Protein Data Bank}}},
  author = {Berman, Helen M. and Westbrook, John and Feng, Zukang and Gilliland, Gary and Bhat, T. N. and Weissig, Helge and Shindyalov, Ilya N. and Bourne, Philip E.},
  year = {2000},
  month = jan,
  journal = {Nucleic Acids Research},
  volume = {28},
  number = {1},
  pages = {235--242},
  issn = {0305-1048},
  doi = {10.1093/nar/28.1.235}
}

@article{burnhamScopusDatabaseReview2006,
  title = {Scopus Database: A Review},
  shorttitle = {Scopus Database},
  author = {Burnham, Judy F.},
  year = {2006},
  month = mar,
  journal = {Biomedical Digital Libraries},
  volume = {3},
  number = {1},
  pages = {1},
  issn = {1742-5581},
  doi = {10.1186/1742-5581-3-1}
}

@article{butlerSoupedupSearchEngines2000,
  title = {Souped-up Search Engines},
  author = {Butler, Declan},
  year = {2000},
  month = may,
  journal = {Nature},
  volume = {405},
  number = {6783},
  pages = {113--114},
  publisher = {{Nature Publishing Group}},
  issn = {1476-4687},
  doi = {10.1038/35012148},
  copyright = {2000 Macmillan Magazines Ltd.},
  langid = {english}
}

@article{chowdharyNaturalLanguageProcessing2020,
  title = {Natural {{Language Processing}}},
  author = {Chowdhary, K. R.},
  year = {2020},
  journal = {Fundamentals of Artificial Intelligence},
  pages = {603--649},
  publisher = {{Springer, New Delhi}},
  doi = {10.1007/978-81-322-3972-7_19},
  langid = {english}
}

@misc{crossrefRESTAPI2020,
  type = {Website},
  title = {{{REST API}}},
  author = {Crossref},
  year = {2020},
  journal = {Crossref},
  copyright = {CC BY 4.0},
  howpublished = {https://www.crossref.org/documentation/retrieve-metadata/rest-api/},
  langid = {english}
}

@article{degelderGeneralizedExpressionSimilarity2001,
  title = {A Generalized Expression for the Similarity of Spectra: Application to Powder Diffraction Pattern Classification},
  shorttitle = {A Generalized Expression for the Similarity of Spectra},
  author = {{de Gelder}, R. and Wehrens, R. and Hageman, J. A.},
  year = {2001},
  journal = {Journal of Computational Chemistry},
  volume = {22},
  number = {3},
  pages = {273--289},
  issn = {1096-987X},
  doi = {10.1002/1096-987X(200102)22:3<273::AID-JCC1001>3.0.CO;2-0},
  langid = {english}
}

@book{dinnebierPowderDiffractionTheory2008c,
  title = {Powder {{Diffraction}}: {{Theory}} and {{Practice}}},
  shorttitle = {Powder {{Diffraction}}},
  editor = {Dinnebier, R E and Billinge, S J L},
  year = {2008},
  publisher = {{Royal Society of Chemistry}},
  address = {{Cambridge}},
  doi = {10.1039/9781847558237},
  isbn = {978-0-85404-231-9},
  langid = {english}
}

@inproceedings{fergusLearningObjectCategories2005,
  title = {Learning Object Categories from {{Google}}'s Image Search},
  booktitle = {Tenth {{IEEE International Conference}} on {{Computer Vision}} ({{ICCV}}'05) {{Volume}} 1},
  author = {Fergus, R. and {Fei-Fei}, L. and Perona, P. and Zisserman, A.},
  year = {2005},
  month = oct,
  volume = {2},
  pages = {1816-1823 Vol. 2},
  issn = {2380-7504},
  doi = {10.1109/ICCV.2005.142}
}

@article{garfieldWhenCite1996,
  title = {When to {{Cite}}},
  author = {Garfield, Eugene},
  year = {1996},
  month = oct,
  journal = {The Library Quarterly},
  volume = {66},
  number = {4},
  pages = {449--458},
  publisher = {{The University of Chicago Press}},
  issn = {0024-2519},
  doi = {10.1086/602912}
}

@article{gates-rectorPowderDiffractionFile2019,
  title = {The {{Powder Diffraction File}}: A Quality Materials Characterization Database},
  shorttitle = {The {{Powder Diffraction File}}},
  author = {{Gates-Rector}, Stacy and Blanton, Thomas},
  year = {2019},
  month = dec,
  journal = {Powder Diffraction},
  volume = {34},
  number = {4},
  pages = {352--360},
  publisher = {{Cambridge University Press}},
  issn = {0885-7156, 1945-7413},
  doi = {10.1017/S0885715619000812},
  langid = {english}
}

@book{gilmoreInternationalTablesCrystallography2019a,
  title = {International {{Tables}} for {{Crystallography}}: {{Powder}} Diffraction},
  shorttitle = {International {{Tables}} for {{Crystallography}}},
  editor = {Gilmore, C. J. and Kaduk, J. A. and Schenk, H.},
  year = {2019},
  month = jul,
  edition = {First},
  volume = {H},
  publisher = {{International Union of Crystallography}},
  address = {{Chester, England}},
  doi = {10.1107/97809553602060000115},
  isbn = {978-1-118-41628-0}
}

@article{ginspargArXiv202011,
  title = {{{ArXiv}} at 20},
  author = {Ginsparg, Paul},
  year = {2011},
  month = aug,
  journal = {Nature},
  volume = {476},
  number = {7359},
  pages = {145--147},
  publisher = {{Nature Publishing Group}},
  issn = {1476-4687},
  doi = {10.1038/476145a},
  copyright = {2011 Nature Publishing Group, a division of Macmillan Publishers Limited. All Rights Reserved.},
  langid = {english}
}

@article{grazulisCrystallographyOpenDatabase2009d,
  title = {Crystallography {{Open Database}} \textendash{} an Open-Access Collection of Crystal Structures},
  author = {Gra{\v z}ulis, S. and Chateigner, D. and Downs, R. T. and Yokochi, A. F. T. and Quir{\'o}s, M. and Lutterotti, L. and Manakova, E. and Butkus, J. and Moeck, P. and Le Bail, A.},
  year = {2009},
  month = aug,
  journal = {Journal of Applied Crystallography},
  volume = {42},
  number = {4},
  pages = {726--729},
  publisher = {{International Union of Crystallography}},
  issn = {0021-8898},
  doi = {10.1107/S0021889809016690},
  copyright = {http://creativecommons.org/licenses/by/2.0/uk},
  langid = {english}
}

@article{groomCambridgeStructuralDatabase2016,
  title = {The {{Cambridge Structural Database}}},
  author = {Groom, C. R. and Bruno, I. J. and Lightfoot, M. P. and Ward, S. C.},
  year = {2016},
  month = apr,
  journal = {Acta Crystallographica Section B: Structural Science, Crystal Engineering and Materials},
  volume = {72},
  number = {2},
  pages = {171--179},
  publisher = {{International Union of Crystallography}},
  issn = {2052-5206},
  doi = {10.1107/S2052520616003954},
  copyright = {http://creativecommons.org/licenses/by/2.0/uk},
  langid = {english}
}

@article{h.e.mgbemereElectricFieldInducedPhaseTransition2017,
  title = {Electric-Field-Induced Phase Transition in {{Mn-Doped}} ({{K0}}.{{48Na0}}.{{48Li0}}.04){{NbO3}} Lead-Free Ceramics},
  author = {Mgbemere, Henry Ekene and Schneider, Gerold A. and Schmitt, Ljubomira Ana and Hinterstein, Jan Manuel},
  year = {2017},
  journal = {Journal of ceramic science and technology},
  volume = {8},
  number = {1},
  pages = {45--52},
  issn = {2190-9385}
}

@book{hallInternationalTablesCrystallography2006,
  title = {International {{Tables}} for {{Crystallography}}: {{Definition}} and Exchange of Crystallographic Data},
  shorttitle = {International {{Tables}} for {{Crystallography}}},
  editor = {Hall, S. R. and McMahon, B. and Fuess, H. and Hahn, Th. and Wondratschek, H. and M{\"u}ller, U. and Shmueli, U. and Prince, E. and Authier, A. and Kopsk{\'y}, V. and Litvin, D. B. and Rossmann, M. G. and Arnold, E. and Hall, S. and McMahon, B.},
  year = {2006},
  month = oct,
  series = {International {{Tables}} for {{Crystallography}}},
  edition = {First},
  volume = {G},
  publisher = {{International Union of Crystallography}},
  address = {{Chester, England}},
  doi = {10.1107/97809553602060000107},
  isbn = {978-1-4020-5411-2 978-1-4020-3138-0}
}

@article{harrisArrayProgrammingNumPy2020,
  title = {Array Programming with {{NumPy}}},
  author = {Harris, Charles R. and Millman, K. Jarrod and {van der Walt}, St{\'e}fan J. and Gommers, Ralf and Virtanen, Pauli and Cournapeau, David and Wieser, Eric and Taylor, Julian and Berg, Sebastian and Smith, Nathaniel J. and Kern, Robert and Picus, Matti and Hoyer, Stephan and {van Kerkwijk}, Marten H. and Brett, Matthew and Haldane, Allan and {del R{\'i}o}, Jaime Fern{\'a}ndez and Wiebe, Mark and Peterson, Pearu and {G{\'e}rard-Marchant}, Pierre and Sheppard, Kevin and Reddy, Tyler and Weckesser, Warren and Abbasi, Hameer and Gohlke, Christoph and Oliphant, Travis E.},
  year = {2020},
  month = sep,
  journal = {Nature},
  volume = {585},
  number = {7825},
  pages = {357--362},
  publisher = {{Nature Publishing Group}},
  issn = {1476-4687},
  doi = {10.1038/s41586-020-2649-2},
  copyright = {2020 The Author(s)},
  langid = {english}
}

@article{hesterValidatingCIFParser2006,
  title = {A Validating {{CIF}} Parser: {{PyCIFRW}}},
  shorttitle = {A Validating {{CIF}} Parser},
  author = {Hester, J. R.},
  year = {2006},
  month = aug,
  journal = {Journal of Applied Crystallography},
  volume = {39},
  number = {4},
  pages = {621--625},
  publisher = {{International Union of Crystallography}},
  issn = {0021-8898},
  doi = {10.1107/S0021889806015627},
  copyright = {Copyright (c) 2006 International Union of Crystallography},
  langid = {english}
}

@article{hunterMatplotlib2DGraphics2007a,
  title = {Matplotlib: {{A 2D Graphics Environment}}},
  shorttitle = {Matplotlib},
  author = {Hunter, John D.},
  year = {2007},
  journal = {Computing in Science \& Engineering},
  volume = {9},
  number = {3},
  pages = {90--95},
  issn = {1521-9615},
  doi = {10.1109/MCSE.2007.55}
}

@article{iturbe-zabaloSymmetrymodeAnalysisPhase2013,
  title = {Symmetry-Mode Analysis of the Phase Transitions in {{SrLaZnRuO}} {\textsubscript{6}} and {{SrLaMgRuO}} {\textsubscript{6}} Ordered Double Perovskites},
  author = {{Iturbe-Zabalo}, E. and Igartua, J. M. and Gateshki, M.},
  year = {2013},
  month = aug,
  journal = {Journal of Applied Crystallography},
  volume = {46},
  number = {4},
  pages = {1085--1093},
  issn = {0021-8898},
  doi = {10.1107/S0021889813013253}
}

@article{jainCommentaryMaterialsProject2013d,
  title = {Commentary: {{The Materials Project}}: {{A}} Materials Genome Approach to Accelerating Materials Innovation},
  shorttitle = {Commentary},
  author = {Jain, Anubhav and Ong, Shyue Ping and Hautier, Geoffroy and Chen, Wei and Richards, William Davidson and Dacek, Stephen and Cholia, Shreyas and Gunter, Dan and Skinner, David and Ceder, Gerbrand and Persson, Kristin A.},
  year = {2013},
  month = jul,
  journal = {APL Materials},
  volume = {1},
  number = {1},
  pages = {011002},
  publisher = {{American Institute of Physics}},
  doi = {10.1063/1.4812323}
}

@article{johnsonBillionScaleSimilaritySearch2021,
  title = {Billion-{{Scale Similarity Search}} with {{GPUs}}},
  author = {Johnson, Jeff and Douze, Matthijs and J{\'e}gou, Herv{\'e}},
  year = {2021},
  month = jul,
  journal = {IEEE Transactions on Big Data},
  volume = {7},
  number = {3},
  pages = {535--547},
  issn = {2332-7790},
  doi = {10.1109/TBDATA.2019.2921572}
}

@article{kasunicStructureLaTi2Al9O19Reanalysis2011,
  title = {Structure of {{LaTi2Al9O19}} and Reanalysis of the Crystal Structure of {{La3Ti5Al15O37}}},
  author = {Kasuni{\v c}, Marta and Meden, Anton and {\v S}kapin, Sre{\v c}o D. and Suvorov, Danilo and Golobi{\v c}, Amalija},
  year = {2011},
  month = dec,
  journal = {Acta Crystallographica. Section B, Structural Science},
  volume = {67},
  number = {Pt 6},
  pages = {455--460},
  issn = {1600-5740},
  doi = {10.1107/S0108768111039759},
  langid = {english},
  pmid = {22101534}
}

@article{mikkiGoogleScholarCompared2009,
  title = {Google {{Scholar}} Compared to {{Web}} of {{Science}}. {{A Literature Review}}},
  author = {Mikki, Susanne},
  year = {2009},
  month = mar,
  journal = {Nordic Journal of Information Literacy in Higher Education},
  volume = {1},
  number = {1},
  issn = {1890-5900},
  doi = {10.15845/noril.v1i1.10},
  langid = {english}
}

@article{mongeonJournalCoverageWeb2016,
  title = {The Journal Coverage of {{Web}} of {{Science}} and {{Scopus}}: A Comparative Analysis},
  shorttitle = {The Journal Coverage of {{Web}} of {{Science}} and {{Scopus}}},
  author = {Mongeon, Philippe and {Paul-Hus}, Ad{\`e}le},
  year = {2016},
  month = jan,
  journal = {Scientometrics},
  volume = {106},
  number = {1},
  pages = {213--228},
  issn = {0138-9130, 1588-2861},
  doi = {10.1007/s11192-015-1765-5},
  langid = {english}
}

@article{orayechModecrystallographyAnalysisCrystal2015,
  title = {Mode-Crystallography Analysis of the Crystal Structures and the Low- and High-Temperature Phase Transitions in {{Na0}}.{{5K0}}.{{5NbO3}}},
  author = {Orayech, B. and Faik, A. and L{\'o}pez, G. A. and Fabelo, O. and Igartua, J. M.},
  year = {2015},
  month = apr,
  journal = {Journal of Applied Crystallography},
  volume = {48},
  number = {2},
  pages = {318--333},
  publisher = {{International Union of Crystallography}},
  issn = {1600-5767},
  doi = {10.1107/S1600576715000941},
  copyright = {Copyright (c) 2015 International Union of Crystallography},
  langid = {english}
}

@article{palaciosPhasesCH34N2003,
  title = {The Phases of [({{CH3}}){{4N}}]({{ClO4}}) at Low Temperature},
  author = {Palacios, El{\'i}as and Burriel, Ram{\'o}n and Ferloni, Paolo},
  year = {2003},
  month = oct,
  journal = {Acta Crystallographica. Section B, Structural Science},
  volume = {59},
  number = {Pt 5},
  pages = {625--633},
  issn = {0108-7681},
  doi = {10.1107/s0108768103019013},
  langid = {english},
  pmid = {14586083}
}

@article{paskinUniqueIdentifiers1999,
  title = {Toward Unique Identifiers},
  author = {Paskin, N.},
  year = {1999},
  month = jul,
  journal = {Proceedings of the IEEE},
  volume = {87},
  number = {7},
  pages = {1208--1227},
  issn = {1558-2256},
  doi = {10.1109/5.771073}
}

@article{samadzadehComparisonFourSearch2013,
  title = {Comparison of {{Four Search Engines}} and Their Efficacy {{With Emphasis}} on {{Literature Research}} in {{Addiction}} ({{Prevention}} and {{Treatment}})},
  author = {Samadzadeh, Gholam Reza and Rigi, Tahereh and Ganjali, Ali Reza},
  year = {2013},
  journal = {International Journal of High Risk Behaviors \& Addiction},
  volume = {1},
  number = {4},
  pages = {166--171},
  issn = {2251-8711},
  doi = {10.5812/ijhrba.6551},
  langid = {english},
  pmcid = {PMC4070130},
  pmid = {24971257}
}

@article{sciauStructuresPhasesParaelectrique1999,
  title = {Structures Des Phases Para\'electrique et Ferro\'electrique de {{Pb2KNb5O15}}},
  author = {Sciau, null and Calvarin, null and Ravez, null},
  year = {1999},
  month = aug,
  journal = {Acta Crystallographica. Section B, Structural Science},
  volume = {55},
  number = {Pt 4},
  pages = {459--466},
  issn = {1600-5740},
  doi = {10.1107/s0108768198017984},
  langid = {english},
  pmid = {10927388}
}

@article{stahliHydrogensubstitutedVtricalciumPhosphate2016,
  title = {Hydrogen-Substituted {$\beta$}-Tricalcium Phosphate Synthesized in Organic Media},
  author = {St{\"a}hli, C. and Th{\"u}ring, J. and Galea, L. and Tadier, S. and Bohner, M. and D{\"o}belin, N.},
  year = {2016},
  month = dec,
  journal = {Acta Crystallographica Section B: Structural Science, Crystal Engineering and Materials},
  volume = {72},
  number = {6},
  pages = {875--884},
  publisher = {{International Union of Crystallography}},
  issn = {2052-5206},
  doi = {10.1107/S2052520616015675},
  copyright = {http://creativecommons.org/licenses/by/2.0/uk},
  langid = {english}
}

@article{strutynskaRietveldRefinementAgCa102013,
  title = {Rietveld Refinement of {{AgCa10}}({{PO4}})7 from {{X-ray}} Powder Data},
  author = {Strutynska, N. Y. and Zatovsky, I. V. and Ogorodnyk, I. V. and Slobodyanik, N. S.},
  year = {2013},
  month = may,
  journal = {Acta Crystallographica Section E: Structure Reports Online},
  volume = {69},
  number = {5},
  pages = {i23-i23},
  publisher = {{International Union of Crystallography}},
  issn = {1600-5368},
  doi = {10.1107/S1600536813007848},
  copyright = {http://creativecommons.org/licenses/by/2.0/uk},
  langid = {english}
}

@incollection{tobyPowderDictionaryPdCIF2006,
  title = {Powder Dictionary ({{pdCIF}})},
  booktitle = {International {{Tables}} for {{Crystallography}}},
  author = {Toby, B. H.},
  editor = {Fuess, H. and Hahn, Th. and Wondratschek, H. and M{\"u}ller, U. and Shmueli, U. and Prince, E. and Authier, A. and Kopsk{\'y}, V. and Litvin, D. B. and Rossmann, M. G. and Arnold, E. and Hall, S. and McMahon, B. and Hall, S. R. and McMahon, B.},
  year = {2006},
  month = oct,
  edition = {First},
  volume = {G},
  pages = {258--269},
  publisher = {{International Union of Crystallography}},
  address = {{Chester, England}},
  doi = {10.1107/97809553602060000742},
  isbn = {978-1-4020-5411-2 978-1-4020-3138-0},
  langid = {english}
}

@article{vannoordenGoogleScholarPioneer2014,
  title = {Google {{Scholar}} Pioneer on Search Engine's Future},
  author = {Van Noorden, Richard},
  year = {2014},
  month = nov,
  journal = {Nature},
  publisher = {{Nature Publishing Group}},
  issn = {1476-4687},
  doi = {10.1038/nature.2014.16269},
  copyright = {2014 Nature Publishing Group},
  langid = {english}
}

@article{virtanenSciPyFundamentalAlgorithms2020b,
  title = {{{SciPy}} 1.0: Fundamental Algorithms for Scientific Computing in {{Python}}},
  shorttitle = {{{SciPy}} 1.0},
  author = {Virtanen, Pauli and Gommers, Ralf and Oliphant, Travis E. and Haberland, Matt and Reddy, Tyler and Cournapeau, David and Burovski, Evgeni and Peterson, Pearu and Weckesser, Warren and Bright, Jonathan and {van der Walt}, St{\'e}fan J. and Brett, Matthew and Wilson, Joshua and Millman, K. Jarrod and Mayorov, Nikolay and Nelson, Andrew R. J. and Jones, Eric and Kern, Robert and Larson, Eric and Carey, C. J. and Polat, {\.I}lhan and Feng, Yu and Moore, Eric W. and VanderPlas, Jake and Laxalde, Denis and Perktold, Josef and Cimrman, Robert and Henriksen, Ian and Quintero, E. A. and Harris, Charles R. and Archibald, Anne M. and Ribeiro, Ant{\^o}nio H. and Pedregosa, Fabian and {van Mulbregt}, Paul},
  year = {2020},
  month = mar,
  journal = {Nature Methods},
  volume = {17},
  number = {3},
  pages = {261--272},
  publisher = {{Nature Publishing Group}},
  issn = {1548-7105},
  doi = {10.1038/s41592-019-0686-2},
  copyright = {2020 The Author(s)},
  langid = {english}
}

@article{vombrockeStandingShouldersGiants2015,
  title = {Standing on the {{Shoulders}} of {{Giants}}: {{Challenges}} and {{Recommendations}} of {{Literature Search}} in {{Information Systems Research}}},
  shorttitle = {Standing on the {{Shoulders}} of {{Giants}}},
  author = {{vom Brocke}, Jan and Simons, Alexander and Riemer, Kai and Niehaves, Bjoern and Plattfaut, Ralf and Cleven, Anne},
  year = {2015},
  month = aug,
  journal = {Communications of the Association for Information Systems},
  volume = {37},
  number = {1},
  issn = {1529-3181},
  doi = {10.17705/1CAIS.03709}
}

@article{zagoracRecentDevelopmentsInorganic2019,
  title = {Recent Developments in the {{Inorganic Crystal Structure Database}}: Theoretical Crystal Structure Data and Related Features},
  shorttitle = {Recent Developments in the {{Inorganic Crystal Structure Database}}},
  author = {Zagorac, D. and M{\"u}ller, H. and Ruehl, S. and Zagorac, J. and Rehme, S.},
  year = {2019},
  month = oct,
  journal = {Journal of Applied Crystallography},
  volume = {52},
  number = {5},
  pages = {918--925},
  publisher = {{International Union of Crystallography}},
  issn = {1600-5767},
  doi = {10.1107/S160057671900997X},
  copyright = {https://creativecommons.org/licenses/by/4.0/},
  langid = {english}
}

@article{zatovskyRietveldRefinementWhitlockiterelated2010,
  title = {Rietveld Refinement of Whitlockite-Related {{K}}(0.8){{Ca}}(9.8){{Fe}}(0.2)({{PO}}(4))(7)},
  author = {Zatovsky, Igor V. and Ogorodnyk, Ivan V. and Strutynska, Nataliya Yu and Slobodyanik, Nikolay S. and Sharkina, Nataliya O.},
  year = {2010},
  month = apr,
  journal = {Acta Crystallographica. Section E, Structure Reports Online},
  volume = {66},
  number = {Pt 5},
  pages = {i41-i42},
  issn = {1600-5368},
  doi = {10.1107/S1600536810014327},
  langid = {english},
  pmcid = {PMC2979107},
  pmid = {21578988}
}

@article{zhangStructuresK005Na02009,
  title = {Structures of {{K0}}.{{05Na0}}.{{95NbO3}} (50\textendash 300 {{K}}) and {{K0}}.{{30Na0}}.{{70NbO3}} (100\textendash 200 {{K}})},
  author = {Zhang, N. and Glazer, A. M. and Baker, D. and Thomas, P. A.},
  year = {2009},
  month = jun,
  journal = {Acta Crystallographica Section B: Structural Science},
  volume = {65},
  number = {3},
  pages = {291--299},
  publisher = {{International Union of Crystallography}},
  issn = {0108-7681},
  doi = {10.1107/S0108768109011057},
  copyright = {Copyright (c) 2009 International Union of Crystallography},
  langid = {english}
}

@article{zvirgzdinsStructureDeterminationThree2014,
  title = {Structure Determination of Three Polymorphs of Xylazine from Laboratory Powder Diffraction Data},
  author = {Zvirgzdins, Alvis and Mishnev, Anatolijs and Actins, Andris},
  year = {2014},
  month = apr,
  journal = {Acta Crystallographica Section B, Structural Science, Crystal Engineering and Materials},
  volume = {70},
  number = {Pt 2},
  pages = {342--346},
  issn = {2052-5206},
  doi = {10.1107/S2052520614001140},
  langid = {english},
  pmid = {24675603}
}


\setcounter{figure}{0}
\setcounter{equation}{0}
\setcounter{table}{0}
\makeatletter
\renewcommand{\fnum@figure}{Fig.~S\thefigure}
\renewcommand{\theequation}{S\arabic{equation}}
\renewcommand{\thetable}{S\arabic{table}}
\makeatother


\end{document}